\documentclass[twocolumn]{aastex63}
\def\actaa{Acta Astronomica}

\usepackage{amsmath}

\hypersetup{linkcolor=red,citecolor=blue,filecolor=green,urlcolor=magenta}

\begin{document}

\shorttitle{Anomalous Cepheid PL \& PW relations}
\shortauthors{Ngeow et al.}

\title{Zwicky Transient Facility and Globular Clusters: The Period-Luminosity and Period-Wesenheit Relations for Anomalous Cepheids Supplemented with Large Magellanic Cloud Sample}

\correspondingauthor{C.-C. Ngeow}
\email{cngeow@astro.ncu.edu.tw}

\author[0000-0001-8771-7554]{Chow-Choong Ngeow}
\affil{Graduate Institute of Astronomy, National Central University, 300 Jhongda Road, 32001 Jhongli, Taiwan}

\author[0000-0001-6147-3360]{Anupam Bhardwaj}
\affil{INAF-Osservatorio astronomico di Capodimonte, Via Moiariello 16, 80131 Napoli, Italy}

\author[0000-0002-3168-0139]{Matthew J. Graham}
\affiliation{Division of Physics, Mathematics, and Astronomy, California Institute of Technology, Pasadena, CA 91125, USA}

\author[0000-0001-5668-3507]{Steven L. Groom}
\affiliation{IPAC, California Institute of Technology, 1200 E. California Blvd, Pasadena, CA 91125, USA}

\author[0000-0002-8532-9395]{Frank J. Masci}
\affiliation{IPAC, California Institute of Technology, 1200 E. California Blvd, Pasadena, CA 91125, USA}

\author[0000-0002-0387-370X]{Reed Riddle}
\affiliation{Caltech Optical Observatories, California Institute of Technology, Pasadena, CA 91125, USA} 

\begin{abstract}
  We present the first $gri$-band period-luminosity (PL) and period-Wesenheit (PW) relations for the fundamental mode anomalous Cepheids. These PL and PW relations were derived from a combined sample of five anomalous Cepheids in globular cluster M92 and the Large Magellanic Cloud, both of which have distance accurate to $\sim1\%$ available from literature. Our $g$-band PL relation is similar to the $B$-band PL relation as reported in previous study. We applied our PL and PW relations to anomalous Cepheids discovered in dwarf galaxy Crater II, and found a larger but consistent distance modulus than the recent measurements based on RR Lyrae. Our calibrations of $gri$-band PL and PW relations, even though less precise due to small number of anomalous Cepheids, will be useful for distance measurements to dwarf galaxies.  
\end{abstract}


\section{Introduction}\label{sec1}

Anomalous Cepheids (hereafter ACep; also known as BLBOO type variable stars) are evolved core-helium burning stars that cross the instability strip with masses of $\sim1$ to $\sim2\ M_\odot$ and pulsation periods in between $\sim 0.5$ and $\sim 2.5$~days. Some of the theoretical work and recent reviews on ACep can be found in \citet{cox1988}, \citet{bono1997}, \citet{marconi2004}, \citet{fiorentino2006}, \citet{sandage2006}, and \citet{monelli2022}. Similar to classical Cepheids and RR Lyrae, ACep can pulsate in fundamental mode or first-overtone mode, and follow period-luminosity (PL) relations. On the PL plane, ACep are fainter than classical Cepheids but brighter than Type II Cepheids at a given period \citep[for example, see Figure 5 in][]{soszynski2015}.

The earliest optical PL(Z, where Z represents metallicity) relations for ACep in $B$- and/or $V$-band were derived by \cite{nemec1988}, \citet{nemec1994}, \citet{bono1997}, \citet{pritzl2002}, and \citet{marconi2004}. Later, ACep PL relations were extended to include other filters, such as $I$- and $K$-band, as well as the period-Wesenheit (PW) relations,\footnote{By construction, Wesenheit index is extinction-free \citep{madore1982,madore1991}.} in \citet{marconi2004}, \citet{ripepi2014}, \citet{soszynski2015}, \citet{groenewegen2017}, and \citet{iwanek2018}. Recently, \citet{ripepi2019} and \citet{ripepi2022} derived the PL and PW relations in the {\it Gaia} filters using ACep located in both the Large and Small Magellanic Cloud (LMC and SMC, respectively). 

\begin{figure*}
  \epsscale{1.1}
  \plottwo{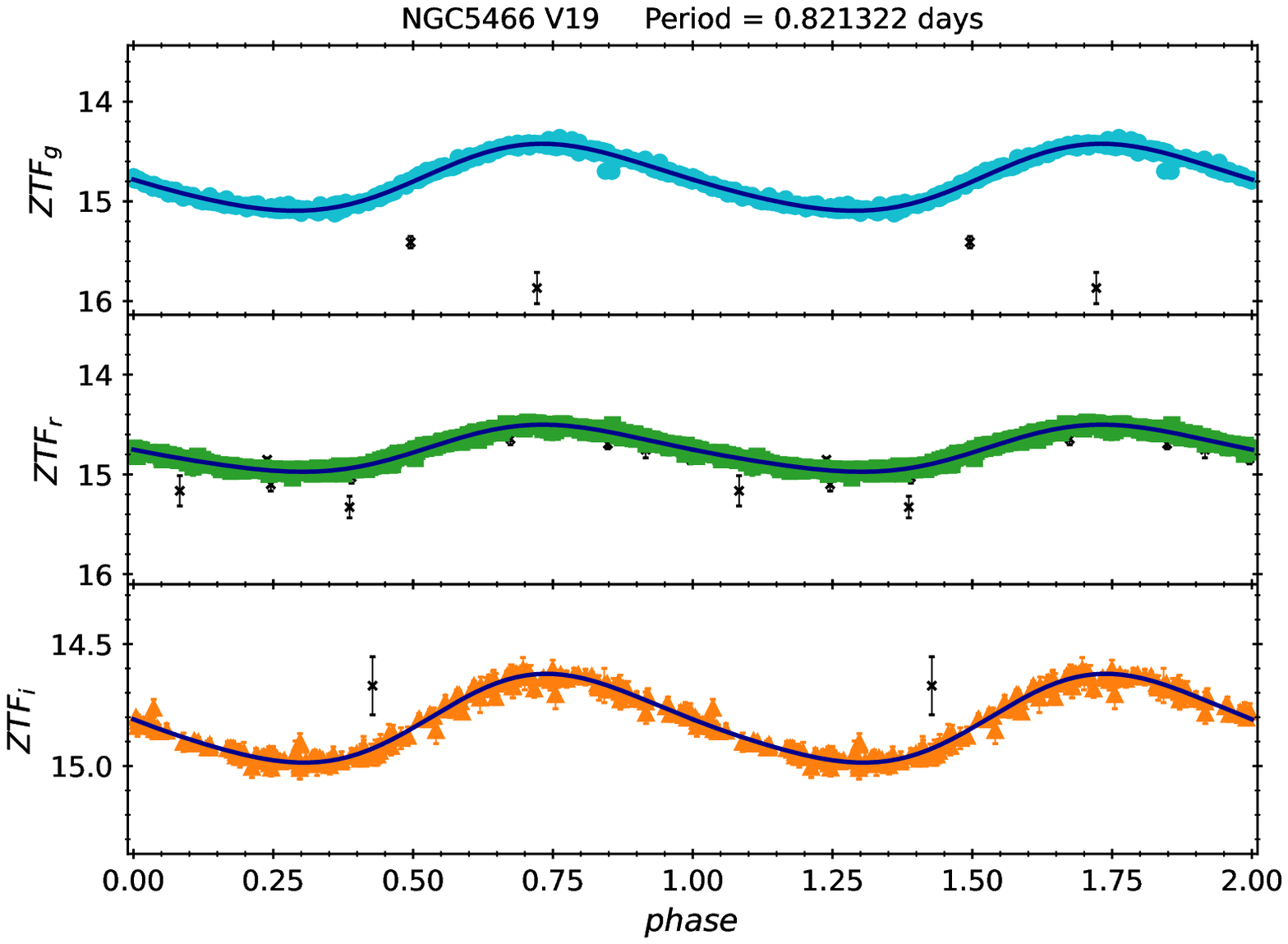}{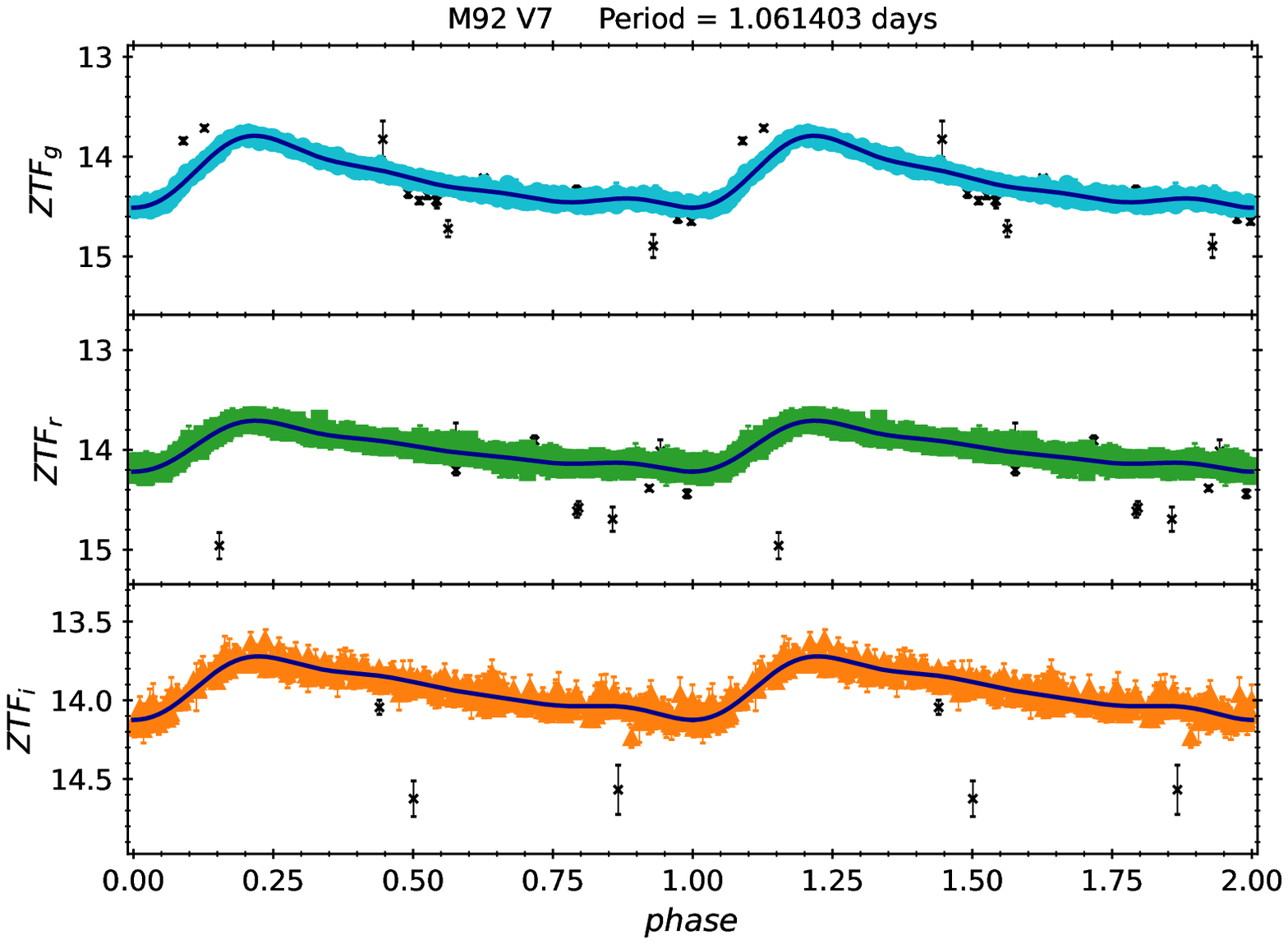}
  \caption{ZTF light curves for the two ACep in GC. The numbers of data points in the $gri$-band light curves are 410, 630, and 180, respectively, for NGC5466 V19, and 1433, 1402, and 432, respectively, for M92 V7. The black crosses are rejected outliers based on a two-step process as described in the text \citep[see also][]{ngeow2022b}. The black curves are the fitted low-order Fourier expansion to the remaining data points on the light curves.}
  \label{fig_lc}
\end{figure*}

To date, empirical ACep PL and PW relations in the optical band are only available in the Johnson-Cousin $BVI$ filters and {\it Gaia} $GB_pR_p$ filters. On the other hand, the Sloan Digital Sky Survey (SDSS) and the SDSS-variant $ugriz$ filters (or a subset of them) are now becoming more popular in a number of time-series synoptic sky surveys, such as the representative Vera C. Rubin Observatory Legacy Survey of Space and Time \citep[LSST,][]{lsst2019}. Similar to our previous work on contact binaries \citep{ngeow2021}, RR Lyrae \citep{ngeow2022a} and Type II Cepheids \citep{ngeow2022b}, we aimed to derive the $gri$-band PL and PW relations for ACep using homogeneous $gri$-band data obtained from the Zwicky Transient Facility \citep[ZTF,][]{bellm2017,bellm2019,gra19,dec20}. ZTF repeatedly observes the northern sky in customized $gri$ filters with a dedicated $47$-squared-degree wide-field mosaic CCD camera mounted on the Palomar 48-inch Samuel Oschin Schmidt telescope. The high-level surveys carried out by ZTF can be divided into public surveys, partner surveys, and California Institute of Technology (Caltech) surveys. All ZTF imaging data, regardless of the surveys, were processed via the same dedicated reduction pipeline \citep{mas19}, and the final catalog products were calibrated to the Pan-STARRS1 \citep[Panoramic Survey Telescope and Rapid Response System 1,][]{chambers2016,magnier2020} AB magnitude system.

ACep are very rare in globular clusters (hereafter GC) but there are a number of known ACep in nearby dwarf galaxies \citep[for example, see][]{nemec1994,marconi2004,monelli2022} including LMC and SMC. Therefore, the empirical PL and PW relations derived in the literature are exclusively based on dwarf galaxies and/or Magellanic Clouds. However, with exceptions of a few cases, dwarf galaxies in general are more distant than GC which limits the accurate calibration of ACep PL and PW relations. The absolute calibration of $gri$-band PL and PW relations for ACep will be useful to obtain independent distances to ACep host dwarf galaxies in the era of LSST. Since there are only a few ACep in the GC (Section \ref{sec2.1}), we also utilized the LMC sample (Section \ref{sec2.2}) to derive PL and PW relations in Section \ref{sec3}. Note that very accurate distances to nearby GCs are now known from \citet{baumgardt2021} and a percent level precise distance to the LMC is available based on late-type eclipsing binaries \citep{pie2019}. We test our relations to a dwarf galaxy in Section \ref{sec4}, followed by discussion and conclusion of this work in Section \ref{sec5}.  

\section{Sample and Data} \label{sec2}

\subsection{GC Samples with ZTF Data} \label{sec2.1}

We searched the literature for ACep in GC that are located north of $-30^\circ$ declination (i.e. within the footprint of ZTF), and identified four such ACep. For M9 V12, even though \citet{af2013} argued it is an ACep, later \citet{soszynski2020} re-classified this cluster variable star as a Type II Cepheid. Hence, we discarded this variable star, which leave three GC ACep in our sample. ZTF light curves for these three ACep were extracted from the PSF (point-spread function) catalogs using a matching radius of $1\arcsec$. These ZTF PSF catalogs include the ZTF partner surveys data until 31 May 2022 and the ZTF Public Data Release 11.\footnote{See \url{https://www.ztf.caltech.edu/ztf-public-releases.html}} 

NGC5466 V19 \citep{zinn1976,mccarthy1997}, a.k.a. BL BOO (the prototype of the BLBOO type variable stars), is a well-known ACep which has been classified as a first-overtone variable star \citep{nemec1988,nemec1994}. We first used the {\tt LombScargleMultiband} module from the {\tt astroML/gatspy}\footnote{\url{https://github.com/astroML/gatspy}, also see \citet{vdp2016}.} package \citep{vdp2015} to search the period of this ACep on the ZTF $gri$-band light curves. A low-order Fourier expansion was used to fit the folded light curves \citep[for more details, see][and reference therein]{ngeow2022b}, and outliers beyond $3s$ from the fitted light curves were identified and removed (where $s$ is the dispersion of the fitted light curve). We then ran the second pass of the {\tt LombScargleMultiband} module to determined the final adopted period, $P=0.821322$~days, for NGC5466 V19. The low-order Fourier expansion was fitted to the folded light curves (with outliers removed) for the second time, determining the following $gri$-band intensity mean magnitudes: $14.748$, $14.738$, and $14.806$~mag, respectively. Due to large number of available data points per light curves (see Figure \ref{fig_lc}), we estimated negligible errors on these mean magnitudes. We note that this two-step process is identical to the one adopted for the Type II Cepheids studied in \citet{ngeow2022b}. The left panel of Figure \ref{fig_lc} presents the folded ZTF light curves for this ACep. 

M92 V7 is another {\it bona fide} ACep located in the GC \citep{osborn2012,yepez2020}, which pulsates in the fundamental mode. Based on the same two-step process as in the case of NGC5466 V19, we determined a period of $1.061403$~days for M92 V7, and the intensity mean magnitudes of $14.214$, $13.989$, and $13.929$~mag in the $gri$-band, respectively. The ZTF light curves for this ACep are shown in the right panel of Figure \ref{fig_lc}.

M15 V142 was re-classified as an ACep in \citet{bhardwaj2021}. However, ZTF light curves for this ACep exhibit strong evidence of blending (i.e. the light curves are ``flat''), therefore we removed this ACep from our sample. 

\subsection{LMC Samples with Archival Data} \label{sec2.2}

\begin{deluxetable*}{lcclllllllr}
  \tabletypesize{\scriptsize}
  \tablecaption{Mean magnitudes and extinction for the five LMC ACep\label{tab_lmc}}
  \tablewidth{0pt}
  \tablehead{
    \colhead{OGLE-IV ID} &
    \colhead{ID in \citet{sebo2002}} &
    \colhead{Period (days)\tablenotemark{a}} &
    \colhead{$B$} &
    \colhead{$V$} &
    \colhead{$I$} &
    \colhead{$g$} &
    \colhead{$r$} &
    \colhead{$i$} &
    \colhead{$E(V-I)$\tablenotemark{b}} &
    \colhead{$\Delta$\tablenotemark{c}}
  }
  \startdata  
  OGLE-LMC-ACEP-019 & Ogle114046 & 0.9094064 & 18.15 & 17.83 & 17.33 & 17.88 & 17.77 & 17.72 & $0.08\pm0.07$ & $\cdots$ \\
  OGLE-LMC-ACEP-021 & Ogle194404 & 1.2958507 & 18.11 & 17.85 & 17.25 & 17.87 & 17.81 & 17.63 & $0.10\pm0.07$ & $-0.00$ \\
  OGLE-LMC-ACEP-026 & Ogle132771 & 1.7387480 & 18.13 & 17.61 & 16.94 & 17.76 & 17.46 & 17.35 & $0.11\pm0.08$ & $+0.03$ \\
  OGLE-LMC-ACEP-046 & Ogle272276 & 1.2637156 & 18.48 & 17.98 & 17.37 & 18.12 & 17.84 & 17.78 & $0.09\pm0.08$ & $+0.03$ \\
  OGLE-LMC-ACEP-050 & Ogle243639 & 1.0446956 & 17.60 & 16.95 & 16.61 & 17.16 & 16.75 & 17.04 & $0.10\pm0.09$ & $-0.12$ \\
  \enddata
  \tablenotetext{a}{Periods adopted from the OGLE-IV LMC ACep catalog \citep{soszynski2015}.}
  \tablenotetext{b}{Extinction retrieved from \citet{sko2021}.}
  \tablenotetext{c}{$\Delta = (V-I)_{SEBO}-(V-I)_{OGLE}$ are the difference of the colors between \citet{sebo2002} and \citet{soszynski2015}.} 
\end{deluxetable*}

\begin{figure*}
  \epsscale{1.1}
  \plottwo{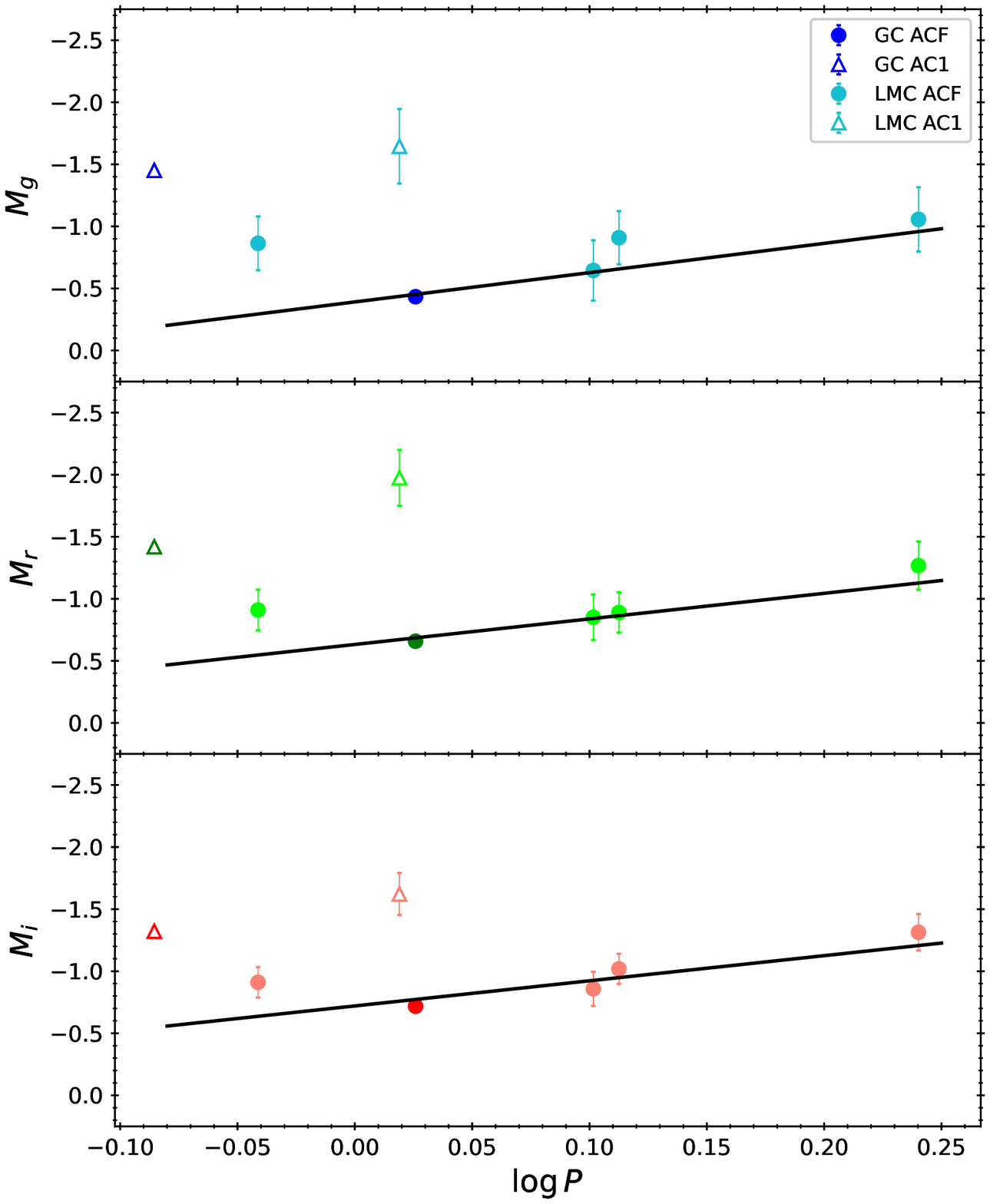}{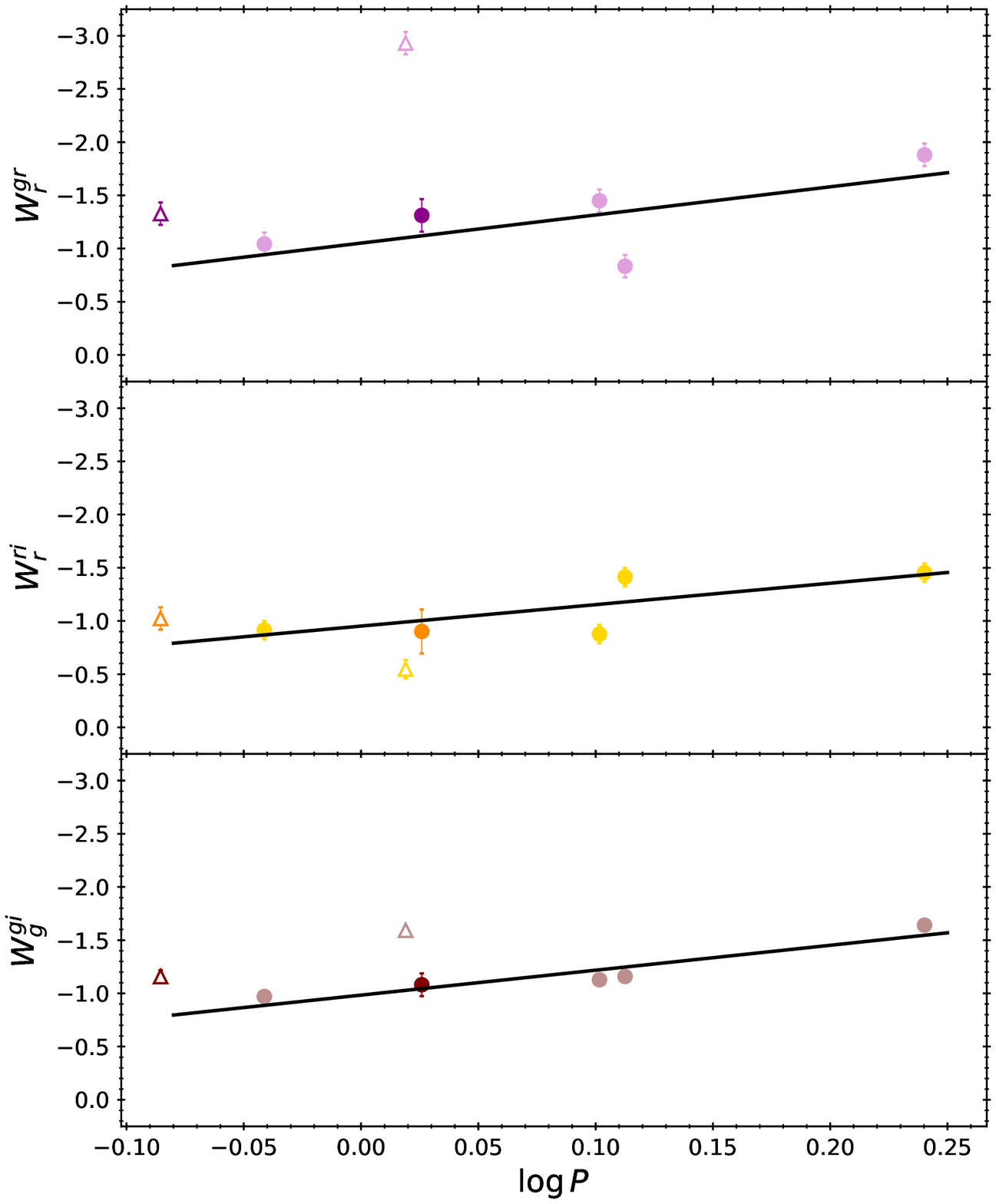}
  \caption{The PL (left panels) and PW (right panels) relations for ACep. Filled circles and open triangles are for the fundamental mode (ACF) and first-overtone (AC1) ACep, while the darker and lighter color symbols represent the GC and LMC ACep, respectively. Solid lines are the fitted PL and PW relations to fundamental mode ACep, as given in equation (1) -- (6).}
  \label{fig_plw}
\end{figure*}

The forth phase of Optical Gravitational Lensing Experiment (OGLE-IV) found 141 ACep in the LMC \citep{soszynski2015}. The $(B-V)$ colors are required to transform the Johnson-Cousin $BVI$-band photometry to the Pan-STARRS1 $gri$-band photometry \citep{tonry2012}. Hence, we cannot use OGLE-IV data which only covers $V$- and $I$-bands. Therefore, we cross-matched the OGLE-IV LMC ACep catalog with the sample of $BVI$ photometry presented in \citet{sebo2002}. In total, we found four common fundamental mode ACep (OGLE-LMC-ACEP-019, 021, 026, and 046) and one first-overtone ACep (OGLE-LMC-ACEP-050). The $BVI$-band mean magnitudes of these five LMC ACep were then transformed to the $gri$-band using the relations provided in \citet{tonry2012}. Both $BVI$-band mean magnitudes and the transformed $gri$-band mean magnitudes for these LMC ACep are provided in Table \ref{tab_lmc}.

\section{The PL and PW Relations} \label{sec3}

Mean magnitudes of the two GC ACep were converted to absolute magnitudes by adopting the GC distances presented in \citet{baumgardt2021}, together with the extinction corrections based on the {\tt Bayerstar2019} 3D reddening map \citep{green2019}.\footnote{See \url{http://argonaut.skymaps.info/usage}, the reddening values were queried via the {\tt dustmaps} package \citep{green2018} available at \url{https://dustmaps.readthedocs.io/en/latest/}.} Similarly, we adopted the most precise LMC distance from \citet{pie2019} and the LMC extinction map of \citet{sko2021} to convert the mean magnitudes of five LMC ACep to absolute magnitudes. We note that accuracy of the adopted distance to NGC5466, M92, and LMC is at the $\sim1\%$~level. As there are only two first-overtone ACep in our sample, we only fit the four LMC and one GC fundamental mode ACep (hereafter collectively referred as calibrating ACep) and obtain the following PL relations:

\begin{figure}
  \epsscale{1.1}
  \plotone{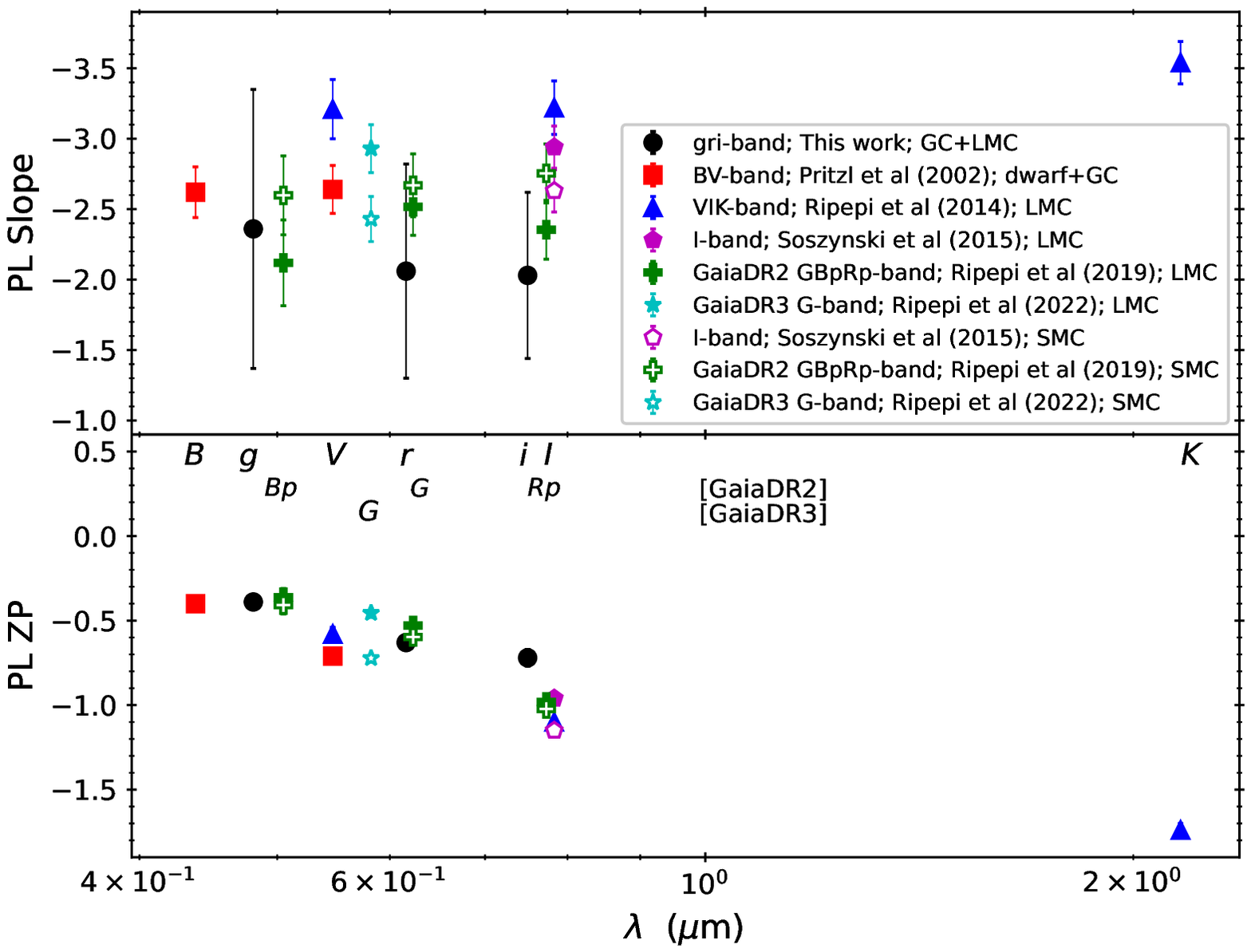}
  \caption{Slopes (upper panel) and zero-points (ZP; lower panel) of the fundamental mode ACep PL relations as a function of wavelengths ($\lambda$). For PL zero-points, we re-scaled the ZP of published LMC and SMC based PL relations with the latest and most precise distance moduli of LMC \citep[][$18.477$~mag]{pie2019} and SMC \citep[][$18.977$~mag]{graczyk2020}. Errors on the PL ZP, in the order of $\sim0.02$ to $\sim 0.05$~mag, are smaller than the size of the symbols. Effective wavelengths for each filters were adopted from the Spanish Virtual Observatory Filter Profile Service \citep{rodrigo2012,rodrigo2020}.}
  \label{fig_compare}
\end{figure}

\begin{eqnarray}
  M_g & = & -2.36[\pm0.99]\log P - 0.39[\pm0.04],\sigma=0.36, \\
  M_r & = & -2.06[\pm0.76]\log P - 0.63[\pm0.04],\sigma=0.23, \\
  M_i & = & -2.03[\pm0.59]\log P - 0.72[\pm0.05],\sigma=0.18,
\end{eqnarray}

\noindent where $\sigma$ is the dispersion of the fitted PL relations. Similar, we obtained the following PW relations \citep[see][]{ngeow2021} based on the calibrating ACep:

\begin{eqnarray}
  W^{gr}_r & = & r - 2.905 (g-r), \nonumber \\
  & = & -2.64[\pm0.51]\log P - 1.05[\pm0.07],\sigma = 0.35, \\
  W^{ri}_r & = & r - 4.051 (r-i), \nonumber \\
  & = & -2.01[\pm0.43]\log P - 0.95[\pm0.06],\sigma = 0.22, \\
  W^{gi}_g & = & g - 2.274 (g-i), \nonumber \\
  & = & -2.34[\pm0.17]\log P - 0.08[\pm0.02], \sigma = 0.11.
\end{eqnarray}

\noindent Figure \ref{fig_plw} presents the $gri$-band PL (left panel) and PW (right panel) relations derived in this work.

Since in general PW relations are expected to exhibit a smaller dispersion than the PL relations, the large dispersions seen in equation (4) and (5) could due to the less accurate colors for some of the LMC ACep derived in \citet{sebo2002}. Last column in Table \ref{tab_lmc} compares the $(V-I)$ colors listed in the OGLE-IV catalog \citep{soszynski2015} and those derived from the mean magnitudes given in \citet{sebo2002}. The color difference for OGLE-LMC-ACEP-050 is $\sim4$~times larger than other ACep listed in Table \ref{tab_lmc}, suggesting the less accurate colors might cause this ACep to be brighter in the $W^{gr}_r$ PW relation but fainter than the fitted relations in the $W^{ri}_r$ PW relations (see the right panels of Figure \ref{fig_plw}, where OGLE-LMC-ACEP-050 is labeled as LMC AC1).

In Figure \ref{fig_compare}, we compared the empirical PL relations, for fundamental mode ACep, in various filters to equation (1) -- (3). In case of the PL slopes, we notice that: (a) in contrast to other pulsating stars (classical Cepheids, Type II Cepheids, and RR Lyrae), there is no clear trends for the PL slopes as a function of wavelengths (i.e. filters), which could due to large uncertainties in the slopes determined in most studies; (b) for a given filter, the PL slopes can be either in agreement (such as in {\it Gaia} filters) or disagreement (such as in $V$-band) among different studies; and (c) PL relations from \citet{ripepi2014} have the steepest slopes, and disagreed with previous works, for the reasons discussed in their study. Despite the large errors due to small number statistics, our $g$-band slopes agree well with the $B$-band PL slope from \citet{pritzl2002}. The PL zero-points (ZP) were found to be more consistent between different studies and follow the expected trend with wavelength. 

\section{An Example of Application: Distance to Crater II} \label{sec4}

\begin{figure}
  \epsscale{1.1}
  \plotone{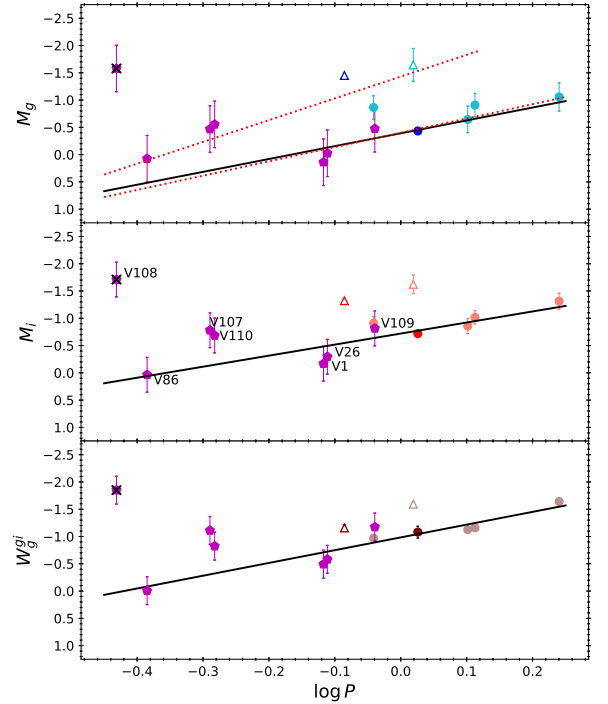}
  \caption{Comparison of the ACep in Crater II (filled magenta pentagons) with the calibrating ACep (the symbols are the same as in Figure \ref{fig_plw}) in the $gi$-band PL and PW relations. Data for ACep in Crater II have been extinction corrected and vertically shifted by a distance modulus of $\mu=20.45$~mag. The filled pentagon with a black cross is V108. The black solid lines are the PL and PW relations derived in Section \ref{sec3}, while the red dotted lines are the $B$-band PL relations (for both fundamental mode and first-overtone ACep) adopted from \citet{pritzl2002}.}
  \label{fig_cr2}
\end{figure}

\citet{vivas2020} reported a discovery of seven ACep in a dwarf galaxy Crater II, providing us an opportunity to test our derived PL and PW relations. As shown in Figure \ref{fig_cr2}, V108 (with the shortest period) is much brighter than the rest of the ACep in Crater II, suggesting this ACep could either be a foreground object or mis-classified as an ACep. Nevertheless, it is clear that V108 should not be used to derive the distance to Crater II. In the top panel of Figure \ref{fig_cr2}, we overlaid the $B$-band PL relations adopted from \citet[][without any photometric transformation]{pritzl2002} together with the $g$-band PL relation derived in equation (1). It is clear that the three longest period ACep (V1, V26, and V109) can be fit well with either the $B$- or $g$-band fundamental mode ACep PL relation, hence they are most likely pulsating in the fundamental mode. Similarly, the three shortest period ACep (V86,\footnote{V86 is located on the $i$-band PL relation and $ W^{gi}_g$ PW relation (see middle and bottom panel of Figure \ref{fig_cr2}) based on the fundamental mode ACep, implying it could be a fundamental mode pulsator. However, as pointed out by the referee, the pulsation period for this ACep is very short and it is unlikely pulsating in the fundamental mode. Unless the PL (and PW) relations for fundamental and first-overtone ACep are in parallel, using locations of the ACep on PL (and PW) plane might not be a reliable method to determine the pulsation mode of ACep, because the un-parallel PL (and PW) relations for both pulsation modes could intersect at short period.} V107 and V110) should be first-overtone pulsators.

The mean magnitudes provided in \citet{vivas2020} are in the SDSS photometric system. Therefore, following the procedures described in \citet{ngeow2022a}, we transformed our $gi$-band PL relations to the SDSS system by adding a small correction term to our PL relations. These correction terms were determined using a subset of calibrating ACep with the same mean $(g-i)_{SDSS}$ colors as the fundamental mode ACep in Crater II. Similarly, we transformed our $W^{gi}_g$ PW relation to the SDSS system and using $W^{gi}_{SDSS}= g_{SDSS}-2.058(g_{SDSS}-i_{SDSS})$. We fit the three longest period fundamental mode ACep with our derived PL and PW relations, and obtained $\mu_g = 20.45\pm0.25$, $\mu_i = 20.55\pm0.29$, and $\mu_W = 20.54\pm0.30$~mag. Errors on these distance moduli $\mu$ were based on the small number statistics \citep[][p. 202]{dean1951,keeping1962} for only three ACep. Extinction corrections were applied to the $g$- and $i$-band mean magnitudes by using the {\tt Bayerstar2019} 3D reddening map \citep{green2019} when deriving the distance moduli for these ACep. 

Using RR Lyrae and a theoretical PL relation, \citet{vivas2020} derived a distance modulus of $20.33\pm0.01$~mag to Crater II. On the other hand, based on the same RR Lyrae sample but with the latest empirical PL relations, \citet{ngeow2022a} argued that the distance modulus of Crater II should be (slightly) larger. Albeit with a larger error (due to small number of samples), the distance modulus derived from ACep is consistent with RR Lyrae-based distance modulus, and supports a farther distance to Crater II.

\section{Discussions and Conclusions} \label{sec5}

Despite the small number of calibrating ACep, we derived for the first time the fundamental mode ACep PL and PW relations in the $gri$-band based on the most precise distances to date to GC and LMC available in the literature. Though the errors on the slopes of our derived PL relations are large, our PL relations fairly agree with published PL relations in other optical filters, especially our $g$-band PL relation is similar to the $B$-band PL relation. Comparison of the PL slopes from various studies also revealed that these PL slopes did not ``converge'' to a trend, implying more data are needed to improve the ACep PL relations in the future. We applied our PL and PW relations to ACep found in dwarf galaxy Crater II, and found that half of the ACep in Crater II are pulsating in fundamental mode. Based on these fundamental mode ACep, we derived a distance to Crater II which agrees with those reported in the literature.

\citet{monelli2022} compiled a list of nearby dwarf galaxies with number of pulsating stars found in these galaxies. As can be seen from their list, almost all nearby dwarf galaxies host at least one RR Lyrae, and about half of them also host ACep. Certainly, for a dwarf galaxy with both RR Lyrae and ACep, RR Lyrae is the preferred distance indicator because in general RR Lyrae are more abundant than ACep and the RR Lyre PL and PW relations are well-developed. On the other hand, ACep are in general brighter, by $\sim 0.4$ to $\sim 2.5$~mag above the horizontal branch \citep[][and reference therein]{vivas2020}, than RR Lyrae because they are more massive. This implies that newly discovered but distant dwarf galaxies near the detection limit of a synoptic sky survey (such as LSST) could miss RR Lyrae but still detecting ACep. In such cases, ACep can provide an independent distance to these dwarf galaxies, and a test for cross-validation for other distance measurements such as using the tip of the red-giant branch (TRGB) method.

\acknowledgments

We are thankful for the useful discussions and comments from an anonymous referee that improved the manuscript. We are thankful for funding from the Ministry of Science and Technology (Taiwan) under the contracts 107-2119-M-008-014-MY2, 107-2119-M-008-012, 108-2628-M-007-005-RSP and 109-2112-M-008-014-MY3.

Based on observations obtained with the Samuel Oschin Telescope 48-inch Telescope at the Palomar Observatory as part of the Zwicky Transient Facility project. ZTF is supported by the National Science Foundation under Grants No. AST-1440341 and AST-2034437 and a collaboration including current partners Caltech, IPAC, the Weizmann Institute of Science, the Oskar Klein Center at Stockholm University, the University of Maryland, Deutsches Elektronen-Synchrotron and Humboldt University, the TANGO Consortium of Taiwan, the University of Wisconsin at Milwaukee, Trinity College Dublin, Lawrence Livermore National Laboratories, IN2P3, University of Warwick, Ruhr University Bochum, Northwestern University and former partners the University of Washington, Los Alamos National Laboratories, and Lawrence Berkeley National Laboratories. Operations are conducted by COO, IPAC, and UW.

This research has made use of the SIMBAD database and the VizieR catalogue access tool, operated at CDS, Strasbourg, France. This research made use of Astropy,\footnote{\url{http://www.astropy.org}} a community-developed core Python package for Astronomy \citep{astropy2013, astropy2018}. This research has made use of the Spanish Virtual Observatory\footnote{\url{https://svo.cab.inta-csic.es}} project funded by MCIN/AEI/10.13039/501100011033/ through grant PID2020-112949GB-I00.

\facility{PO:1.2m}

\software{{\tt astropy} \citep{astropy2013,astropy2018}, {\tt dustmaps} \citep{green2018}, {\tt gatspy} \citep{vdp2015}, {\tt Matplotlib} \citep{hunter2007},  {\tt NumPy} \citep{harris2020}, {\tt SciPy} \citep{virtanen2020}.}




\begin{thebibliography}{} 

\bibitem[Arellano Ferro et al.(2013)]{af2013} Arellano Ferro, A., Bramich, D.~M., Figuera Jaimes, R., et al.\ 2013, \mnras, 434, 1220
  
\bibitem[Astropy Collaboration et al.(2013)]{astropy2013} Astropy Collaboration, Robitaille, T.~P., Tollerud, E.~J., et al.\ 2013, \aap, 558, A33

\bibitem[Astropy Collaboration et al.(2018)]{astropy2018} Astropy Collaboration, Price-Whelan, A.~M., Sip{\H{o}}cz, B.~M., et al.\ 2018, \aj, 156, 123

\bibitem[Baumgardt \& Vasiliev(2021)]{baumgardt2021} Baumgardt, H. \& Vasiliev, E.\ 2021, \mnras, 505, 5957

\bibitem[Bhardwaj et al.(2021)]{bhardwaj2021} Bhardwaj, A., Rejkuba, M., Sloan, G.~C., et al.\ 2021, \apj, 922, 20
  
\bibitem[Bellm \& Kulkarni(2017)]{bellm2017} Bellm, E. \& Kulkarni, S.\ 2017, Nature Astronomy, 1, 0071
  
\bibitem[Bellm et al.(2019)]{bellm2019} Bellm, E.~C., Kulkarni, S.~R., Graham, M.~J., et al.\ 2019, \pasp, 131, 018002

\bibitem[Bono et al.(1997)]{bono1997} Bono, G., Caputo, F., Santolamazza, P., et al.\ 1997, \aj, 113, 2209

\bibitem[Chambers et al.(2016)]{chambers2016} Chambers, K.~C., Magnier, E.~A., Metcalfe, N., et al.\ 2016, arXiv:1612.05560

\bibitem[Cox \& Proffitt(1988)]{cox1988} Cox, A.~N. \& Proffitt, C.~R.\ 1988, \apj, 324, 1042
  
\bibitem[Dean \& Dixon(1951)]{dean1951} Dean, R.~B. \& Dixon, W.~J.\ 1951, Anal. Chem., 23, 636
  
\bibitem[Dekany et al.(2020)]{dec20} Dekany, R., Smith, R.~M., Riddle, R., et al.\ 2020, \pasp, 132, 038001

\bibitem[Fiorentino et al.(2006)]{fiorentino2006} Fiorentino, G., Limongi, M., Caputo, F., et al.\ 2006, \aap, 460, 155

\bibitem[Graczyk et al.(2020)]{graczyk2020} Graczyk, D., Pietrzy{\'n}ski, G., Thompson, I.~B., et al.\ 2020, \apj, 904, 13
  
\bibitem[Graham et al.(2019)]{gra19} Graham, M.~J., Kulkarni, S.~R., Bellm, E.~C., et al.\ 2019, \pasp, 131, 078001

\bibitem[Green(2018)]{green2018} Green, G.~M.\ 2018, The Journal of Open Source Software, 3, 695
  
\bibitem[Green et al.(2019)]{green2019} Green, G.~M., Schlafly, E., Zucker, C., et al.\ 2019, \apj, 887, 93

\bibitem[Groenewegen \& Jurkovic(2017)]{groenewegen2017} Groenewegen, M.~A.~T. \& Jurkovic, M.~I.\ 2017, \aap, 604, A29

\bibitem[Harris et al.(2020)]{harris2020} Harris, C.~R., Millman, K.~J., van der Walt, S.~J., et al.\ 2020, \nat, 585, 357

\bibitem[Hunter(2007)]{hunter2007} Hunter, J.~D.\ 2007, Computing in Science and Engineering, 9, 90

\bibitem[Ivezi{\'c} et al.(2019)]{lsst2019} Ivezi{\'c}, {\v{Z}}., Kahn, S.~M., Tyson, J.~A., et al.\ 2019, \apj, 873, 111

\bibitem[Iwanek et al.(2018)]{iwanek2018} Iwanek, P., Soszy{\'n}ski, I., Skowron, D., et al.\ 2018, \actaa, 68, 213
  
\bibitem[Keeping(1962)]{keeping1962} Keeping, E. S. 1962, Introduction to Statistical Inference (Princeton, NJ: Van Nostrand-Reinhold)

\bibitem[Madore(1982)]{madore1982} Madore, B.~F.\ 1982, \apj, 253, 575

\bibitem[Madore \& Freedman(1991)]{madore1991} Madore, B.~F. \& Freedman, W.~L.\ 1991, \pasp, 103, 933
  
\bibitem[Magnier et al.(2020)]{magnier2020} Magnier, E.~A., Sweeney, W.~E., Chambers, K.~C., et al.\ 2020, \apjs, 251, 5

\bibitem[Marconi et al.(2004)]{marconi2004} Marconi, M., Fiorentino, G., \& Caputo, F.\ 2004, \aap, 417, 1101
  
\bibitem[Masci et al.(2019)]{mas19} Masci, F.~J., Laher, R.~R., Rusholme, B., et al.\ 2019, \pasp, 131, 018003

\bibitem[McCarthy \& Nemec(1997)]{mccarthy1997} McCarthy, J.~K. \& Nemec, J.~M.\ 1997, \apj, 482, 203

\bibitem[Monelli \& Fiorentino(2022)]{monelli2022} Monelli, M. \& Fiorentino, G.\ 2022, Universe, 8, 191
  
\bibitem[Nemec et al.(1988)]{nemec1988} Nemec, J.~M., Wehlau, A., \& Mendes de Oliveira, C.\ 1988, \aj, 96, 528

\bibitem[Nemec et al.(1994)]{nemec1994} Nemec, J.~M., Nemec, A.~F.~L., \& Lutz, T.~E.\ 1994, \aj, 108, 222

\bibitem[Ngeow et al.(2021)]{ngeow2021} Ngeow, C.-C., Liao, S.-H., Bellm, E.~C., et al.\ 2021, \aj, 162, 63
  
\bibitem[Ngeow et al.(2022a)]{ngeow2022a} Ngeow, C.-C., Bhardwaj, A., Dekany, R. et al.\ 2022a, \aj, 163, 239

\bibitem[Ngeow et al.(2022b)]{ngeow2022b} Ngeow, C.-C., Bhardwaj, A., Henderson, J.-Y. et al.\ 2022b, arXiv:2208.03404
  
\bibitem[Osborn et al.(2012)]{osborn2012} Osborn, W., Kopacki, G., \& Haberstroh, J.\ 2012, \actaa, 62, 377

\bibitem[Pietrzy{\'n}ski et al.(2019)]{pie2019} Pietrzy{\'n}ski, G., Graczyk, D., Gallenne, A., et al.\ 2019, \nat, 567, 200

\bibitem[Pritzl et al.(2002)]{pritzl2002} Pritzl, B.~J., Armandroff, T.~E., Jacoby, G.~H., et al.\ 2002, \aj, 124, 1464

\bibitem[Ripepi et al.(2014)]{ripepi2014} Ripepi, V., Marconi, M., Moretti, M.~I., et al.\ 2014, \mnras, 437, 2307

\bibitem[Ripepi et al.(2019)]{ripepi2019} Ripepi, V., Molinaro, R., Musella, I., et al.\ 2019, \aap, 625, A14

\bibitem[Ripepi et al.(2022)]{ripepi2022} Ripepi, V., Clementini, G., Molinaro, R., et al.\ 2022, arXiv:2206.06212

\bibitem[Rodrigo \& Solano(2020)]{rodrigo2020} Rodrigo, C. \& Solano, E.\ 2020, in Contributions to the XIV.0 Scientific Meeting (virtual) of the Spanish Astronomical Society (Barcelona: Spanish Astronomical Society), 182
  
\bibitem[Rodrigo et al.(2012)]{rodrigo2012} Rodrigo, C., Solano, E., \& Bayo, A.\ 2012, IVOA Working Draft
  
\bibitem[Sandage \& Tammann(2006)]{sandage2006} Sandage, A. \& Tammann, G.~A.\ 2006, \araa, 44, 93
  
\bibitem[Sebo et al.(2002)]{sebo2002} Sebo, K.~M., Rawson, D., Mould, J., et al.\ 2002, \apjs, 142, 71

\bibitem[Skowron et al.(2021)]{sko2021} Skowron, D.~M., Skowron, J., Udalski, A., et al.\ 2021, \apjs, 252, 23
  
\bibitem[Soszy{\'n}ski et al.(2015)]{soszynski2015} Soszy{\'n}ski, I., Udalski, A., Szyma{\'n}ski, M.~K., et al.\ 2015, \actaa, 65, 233

\bibitem[Soszy{\'n}ski et al.(2020)]{soszynski2020} Soszy{\'n}ski, I., Udalski, A., Szyma{\'n}ski, M.~K., et al.\ 2020, \actaa, 70, 101

\bibitem[Tonry et al.(2012)]{tonry2012} Tonry, J.~L., Stubbs, C.~W., Lykke, K.~R., et al.\ 2012, \apj, 750, 99
    
\bibitem[VanderPlas \& Ivezi{\'c}(2015)]{vdp2015} VanderPlas, J.~T., \& Ivezi{\'c}, {\v{Z}}.\ 2015, \apj, 812, 18

\bibitem[VanderPlas(2016)]{vdp2016} VanderPlas, J.\ 2016, gatspy: General tools for Astronomical Time Series in Python, ascl:1610.007

\bibitem[Virtanen et al.(2020)]{virtanen2020} Virtanen, P., Gommers, R., Oliphant, T.~E., et al.\ 2020, Nature Methods, 17, 261

\bibitem[Vivas et al.(2020)]{vivas2020} Vivas, A.~K., Walker, A.~R., Mart{\'\i}nez-V{\'a}zquez, C.~E., et al.\ 2020, \mnras, 492, 1061
  
\bibitem[Yepez et al.(2020)]{yepez2020} Yepez, M.~A., Arellano Ferro, A., \& Deras, D.\ 2020, \mnras, 494, 3212

\bibitem[Zinn \& Dahn(1976)]{zinn1976} Zinn, R. \& Dahn, C.~C.\ 1976, \aj, 81, 527
  
\end{thebibliography}
\end{document}